\documentclass[12pt]{article}
\usepackage{amsfonts}
\usepackage{epsfig}
\newcommand{\nn}{\nonumber}

\newcommand{\be}{\begin{equation}}
\newcommand{\ee}{\end{equation}}
\newcommand{\bd}{\begin{displaymath}}
\newcommand{\ed}{\end{displaymath}}
\newcommand{\bea}{\begin{eqnarray}}
\newcommand{\eea}{\end{eqnarray}}
\newcommand{\eps}{\epsilon}
\renewcommand{\paragraph}[1]{
\vspace{.8mm}\par\noindent {\sl #1}\\
\vspace{0.2mm} }

\newcommand{\ba}{\left(\begin{array}}
\newcommand{\ea}{\end{array}\right)}
\newcommand{\Z}{{\cal Z}}
\def\codim{{\rm codim}\ }
\def\dim{{\rm dim}\ }

\def\L{{\cal L}}
\def\C{{\cal C}}

\begin{document}
\begin{titlepage}
\begin{flushright}
SU-ITP-00/01\\
{\tt hep-th/0001072}\\
\today\\
\end{flushright}
\vskip 2cm

\begin{center}
{\large \textsc{ Twistors and Actions on Coset Manifolds}} \vskip 0.7 cm
{\small
{\bf  Yonatan Zunger}\\
\vskip.5cm
{\it Department of Physics, \\
Stanford University, Stanford, CA 94305-4060, USA \\
E-mail: zunger@leland.stanford.edu}
}
\end{center}
\vskip 1cm
\begin{abstract}
Particle and string actions on coset spaces typically lack a quadratic
kinetic term, making their quantization difficult. We define a notion
of twistors on these spaces, which are hypersurfaces in a vector space
that transform linearly under the isometry group of the coset. By
associating the points of the coset space with these hypersurfaces, and the 
internal coordinates of these hypersurfaces with momenta, it is possible
to construct manifestly symmetric actions with leading quadratic terms.
We give a general algorithm and work out the case of a particle on 
$AdS_p$ explicitly. In this case, the resulting action is a world-line
gauge theory with sources, (the gauge group depending on $p$) which is
equivalent to a nonlocal world-line $\sigma$-model. 
\end{abstract}

%\vfill

%\footnoterule
%\noindent
%{\footnotesize
%$\phantom{a}^a$ e-mail: zunger@leland.stanford.edu. }
\end{titlepage}

\section{Introduction}

The standard action for a particle or string on a coset space $G/H$ is 
manifestly invariant under $G$ but does not have a quadratic kinetic term. 
This obstructs the usual quantization procedure. Moreover, the isometries 
are nonlinearly realized on the coordinates and so even if the action
were quadratic, the fields would not automatically form $G$-representations.
This makes it difficult to directly study systems such as strings on 
$AdS_{p+2}\times S^{d-p-2}$ which are important for understanding 
holography. \cite{Malda,GubKleb,Witten}

 A hint on how to work around this comes from twistors. These were originally
set up by Penrose to study conformal Minkowski space, \cite{Penrose} and 
have since been generalized to conformal superspace \cite{Ferber} and 
$AdS_5$. \cite{CRZ, CKR} In all of these cases,
twistors associated a hypersurface in some vector space which transforms
linearly under the isometry group with every point of the coset space.
The internal coordinates of these hypersurfaces were associated with
momenta and constrained quantities. 

 This construction can be generalized to arbitrary coset 
manifolds\footnote{The results below apply both to cosets and supercosets,
with no additional restrictions (reductivity, symmetry, etc.) except 
where explicitly noted.}
$G/H$. A mapping between points of the coset and hypersurfaces in a 
vector space can be constructed which naturally mimics the geometric structure
of the coset. The isometries, for example, can easily be extracted from the
linear isometry transformations of the twistors. 

 If the vector space is also a Hilbert space (i.e. posesses an appropriate 
inner product) then one can naturally construct objects out of twistors 
which are manifestly invariant under the coset isometries. Using the twistor
mapping, these can be written as (typically fairly complicated) functions
of the coset coordinates and the internal coordinates of the twistor. 
Since these quantities are manifestly $G$-invariant, one can construct
actions out of them which are equivalent to ordinary coset actions if
the internal coordinates are identified with momenta. Since the twistor
mapping is typically very complicated, very simple twistor actions are
equivalent to very complicated coset actions. 

 We demonstrate this construction for a particle on $AdS_p$. Twistors are
built in a vector space which transforms in the spinor representation
of $SO(p-1,2)$. A world-line gauge theory can be built out of these
twistors which is equivalent to the ordinary action for a massive
particle on the coset. This theory is equivalent to a nonlocal 
$\sigma$-model whose target space is the vector space. This construction
can probably be generalized to the study of particles and strings on
anti-de Sitter superspaces such as those important for the AdS/CFT 
correspondence.

\section{The twistor construction}

We begin by describing cosets in a language which naturally leads to
twistors.  A point in a coset $G/H$ is associated with a hypersurface 
in the group manifold by the relation
\be
\phi(\hat x\in G) = \left\{\hat xh: h\in H\right\}\ .
\label{eq:coset1}
\ee
The $\hat x$ which generate distinct $\phi(\hat x)$ are given by
\be
\hat x := v(\xi) \hat x_0\ ,
\label{eq:coset2}
\ee
where $\hat x_0\in G$ is the origin of the coset space, (an arbitrary
point) and $\xi$ is a collective coordinate on $G/H$. The function 
$v(\xi)$ is a coset representative, which for our purposes is a function
from the coordinates to the group such that $v(0)=1$ and $\phi\circ v$
is 1-1, so distinct hypersurfaces are associated with distinct coordinates.
A particular form of $v(\xi)$ which we will often use is
\be
v(\xi) = e^{\xi\cdot K} h(\xi)\ ,
\ee
where the $K$ are the generators of $G$ not in $H$, and $h(\xi)$ is some
$H$-valued function chosen to simplify the resulting expression.

Using this function, we can associate coset coordinates with hypersurfaces in the
group manifold in a natural way:
\be
\phi(\xi) = \phi(\hat{x}(\xi)) = v(\xi) \phi(\hat x_0)\ .
\ee
These hypersurfaces are naturally invariant under the right action of $H$.

Such a construction cannot always be globally performed. The problem 
is analogous to the selection of coordinates on a sphere $S^2=SO(3)/SO(2)$.
Technically, it arises because the coset representative $v(x)$ is a 
section of the principal bundle $G\rightarrow G/H$ and so in general 
cannot be globally defined. We may resolve this issue analogously to the
problem of coordinates by covering $G/H$ with patches (open sets whose
intersections are contractible) and performing this construction on each
open set. Transition functions on intersections are naturally induced
from the transition functions of the principal $G$-bundle of which $v$
is a section. It is also necessary to choose a different $\hat{x}_0$
for each patch, which is analogous to using the north and south poles as
origins of the two coordinate systems on $S^2$. The result of such a 
construction will (as we will see below) be a twistor bundle rather than
a global twistor space, but this will not introduce any unfamiliar
complexities.

Using this construction, the geometry on the coset space may be defined
by the invariances of the Cartan form $L=v^{-1}dv$. This form is 
canonically separated into $L=E\cdot K + \Omega\cdot H$, where $E$ is
the vielbein and $\Omega$ the $H$-connection. (This generalizes
the spin connection of Minkowski space) When $G$ is semisimple, 
this can be contracted with a restricted Cartan-Killing metric to give a 
metric on the coset;
\be
g^{\mu\nu} = \eta^{AB} L^\mu_A L^\nu_B\ ,
\ee
where the indices $A$ and $B$ run over only the coset ($K$) generators of
$G$.\footnote{This procedure is discussed in detail in \cite{BlueBook}.}
In the more general case (which includes Minkowski space) the procedure
is somewhat more subtle. For some groups at least, there is an invariant
symmetric two-form which may take the place of the Cartan-Killing metric
$\eta$; however, there is no general existence proof nor is there a 
method of computing such forms. In this case one assumes that the 
transformations which lead to covariant transformations of the $K$-components
of the Cartan form will become isometries of the coset spaces.

The isometries of this space are given implicitly by the action of
$G$ on the coset representative:
\be
\delta \xi: \delta v(\xi) = gv(\xi)\ , 
\label{eq:cosetisom}
\ee
where $g\in G$. This implies that
\be
\delta\phi(\xi) = g\phi(\xi)\ .
\label{eq:vareqn}
\ee

It is straightforward to compute the actual transformations of the 
$\xi$ from this relationship if we write (\ref{eq:vareqn}) in an
explicit representation. This motivates us to define 
twistors to be explicit group representations
of these hypersurfaces $\phi(\xi)$. Specifically, if we represent the
group on a Hilbert space $\Lambda$, (not necessarily infinite-dimensional!)
a twistor is a mapping of coordinates to $H$-invariant hypersurfaces
in $\Lambda$ given by
\be
\Z(\xi)=v(\xi)\Z_0\ ,
\ee
where $v(\xi)$ is the $\Lambda$-representation of the coset representative, and $\Z_0$ 
is an $H$-invariant hypersurface in $\Lambda$. (The $\Lambda$-representation of
$\phi(\hat{x}_0)$) As before, the mapping $\Z$ must be 1-1 for
the set of $\Z(\xi)$ to be isomorphic to the coset, which means that the
codimension of $\Z_0$ in $\Lambda$ must be no less than the dimension of the 
coset. ($\codim\Z(\xi)=\codim\Z_0$ for any $\xi$ since $v(x)$ is surjective) 
We will also restrict ourselves to $\dim\Z_0>0$, since otherwise $\Z$
would be a mapping of points onto points and so would lose several of
the interesting features which we will discover below.

We next write $\Z_0$ explicitly as a linear function of some coordinates
on $\Lambda$. (These coordinates may be curvilinear, although we do not
consider such possibilities in depth here) Then using the explicit form
of the coset representative $v(\xi)$, we may write each $\Z(\xi)$ as a 
function of the coordinates $\xi$ and the internal coordinates $\lambda$
of $\Z_0$ which is linear in $\lambda$ and typically fairly complicated
in $\xi$. 

This process has two advantages.  First, since the $\Z$ are given as
explicit functions of the coset coordinates, it is straightforward to use
(\ref{eq:cosetisom}) to compute the geometric properties of the space.
This is especially valuable in the case of complicated cosets such as
$AdS_p\times S^{d-p}$ superspace, where traditional (differential-equation)
methods of calculating isometries are very difficult. Second, since
$\Lambda$ is a Hilbert space there is a natural continuous and complete
inner product of twistors which is manifestly invariant under the action
of $G$. This allows one to easily construct quantities with a very
complicated dependence on $\xi$ and $\lambda$ which are invariant under
$G$. This invariance persists even though $H\subset G$ is typically 
nonlinearly realized on the coset space. If (as we will do later) we
identify the $\lambda$ with some internal parameters of a system such
as momenta, it is possible to use these invariants to
construct very simple twistor-based actions which are equivalent to very
complicated coordinate-space actions.

\smallskip

As the preceding discussion was somewhat abstract, it is useful to 
consider some explicit examples. We begin with the case of conformal
Minkowski space $SO(3,2)/ISO(3,1)\times D$, where $D$ is the dilatation 
operator. We choose as our representation the 4-component spinor 
representation of $SO(3,2)$, which decomposes into a 2-component spinor
and a 2-component conjugate spinor of $ISO(3,1)$ of conformal weights
$\pm 1/2$.\footnote{Although we use spinorial representations here,
this is by no means a general feature of twistors.} In this representation,
a group element has the form
\be
g=\left(\begin{array}{cc} L_\beta{}^\alpha + \frac{1}{2}D\delta_\beta{}^\alpha
& -i K_{\alpha\dot\alpha} \\ -i P^{\dot\alpha\alpha} & -\bar{L}^{\dot\alpha}
{}_{\dot\beta}-\frac{1}{2}\delta^{\dot\alpha}{}_{\dot\beta} \end{array}
\right)
\ee
and the initial hypersurface $\Z_0$ has the form
\be
\Z_0 = \left(\begin{array}{c}\lambda_\alpha\\ \mu^{\dot\alpha}\end{array}\right)\ ,
\ee
where the $\alpha$ ($\dot\alpha$) are (conjugate) spinor indices of $SO(3,1)$
and $\lambda$ and $\mu$ are complex. The stability group $H$ is generated
by the $L$, $K$, and $D$. Lorentz invariance implies that if $\Z_0$ has any 
point with $\lambda\not=0$, 
it must contain all such points, and likewise for $\mu$. Thus the
dimension constraint $0<\dim\Z_0\le 4$ requires that exactly one
of the two be independent. Without loss of generality, we choose $\lambda$
and let $\mu$ be a linear function thereof. 
The remaining part of $H$-invariance then determines $\mu=0$. 

Now let us choose the canonical coset representative
\be
v(x) = e^{-ix\cdot P} = \left(\begin{array}{cc}1&0\\-ix^{\dot\alpha\alpha} &
1\end{array}\right)\ .
\ee
The twistor mapping is now
\be
\Z(x) = \left(\begin{array}{c}\lambda_\alpha \\ -ix^{\dot\alpha\alpha}
\lambda_\alpha\end{array}\right)\ .
\label{eq:twmap}
\ee
This is the familiar Penrose twistor formula. \cite{Penrose} We will not
discuss isometries and invariants in this case, saving that instead for
the more detailed example of $AdS_p$ below.

It is worth noting that this procedure was by no means
unique. The freedoms of choice are in the selection of an appropriate
coset representative (which will typically be determined by algebraic
simplicity, subject to the requirement that $\Z(\xi)$ is 1-1) and in the
choice of the initial hypersurface $\Z_0$.

We can also naturally ask about the invariants which may be constructed
out of these twistors. The simplest world-line action which one may
construct out of these twistors is clearly
\be
\L = i\bar\Z \partial\Z\ ,
\label{eq:penroseaction}
\ee
where contraction has been performed with the standard spinor metric. If
we substitute in (\ref{eq:twmap}), and write
\be
P_{\dot\alpha\alpha} = \bar\lambda_{\dot\alpha} \lambda_\alpha\ ,
\ee
then this action reduces to the simple form
\be
S=i\int d\tau P\cdot \partial x
\ee
which is the usual world-line action for a massless particle. $P$ is 
automatically null in this case because the spinor metric $\eps^{\alpha\beta}$
is antisymmetric. Massive actions cannot easily be written in terms of
these twistors, which is unsurprising since we are here working in conformal 
Minkowski space. 

In this case we have put the internal coordinates $\lambda_\alpha$ of
the twistor to use as momenta of the particle. It is not clear how
general such an interpretation is; clearly a precondition for the
possibility of so doing is that the twistor bundle (the set of these
hypersurfaces $\Z(x)$ over every point, with open sets as discussed
above) contains the tangent bundle of $G/H$ as a subbundle. Even when
this is not possible, the procedure above will turn the $\lambda_\alpha$
into Lagrange multipliers for various quantities; when the quantities are
derivatives of the coordinates, there is a somewhat natural momentum
interpretation. We will see more of this construction later.

\section{Twistorization of $AdS_p$}

We now turn to the case of particles on $AdS_p=SO(p-1,2)/SO(p-1,1)$. 
The ordinary world-line action for these particles is manifestly 
$G$-invariant but does not have a quadratic kinetic term, so it
is useful to try to rephrase this in terms of twistors. This is 
reasonable since the first-order action,
\be
\L = \frac{1}{2}P\cdot \partial x + P_\rho \partial \rho + u\left[\frac{1}
{2\rho^2} P^2 - \rho^2 P_\rho^2 - m^2 R^2\right]
\label{eq:cosetaction}
\ee
contains only terms of the form $P\cdot \partial x$, which are similar to
those found in the conformal Minkowski action (\ref{eq:penroseaction}), 
and a constraint term which is $G$-invariant although not manifestly so. 
In a twistor construction one
hopes that this can be rewritten in a manifestly symmetric (and preferably
simple) way, and we will see that this is indeed the case.

Twistorization must begin with a choice of $G$-representation. The two
simplest choices are the fundamental and the spinor. The fundamental has 
simpler group
generators, but since its dimension is $(p+1)$ such twistors would have
only one internal coordinate and so momenta could not be encoded by the
twistor. Therefore we use the spinor representation, which has complex 
dimension $2^{\left\lfloor(p+1)/2\right\rfloor}\equiv 2d$. The group elements
in this representation are\footnote{Our index notation is: $\mu=0\ldots p-2$
is an $SO(p-2,1)$ (Lorentz) vector index, $\alpha,\beta$ and
 $\dot\alpha,\dot\beta=1\ldots d$ are Lorentz spinor and conjugate spinor 
indices, respectively. $A,B$ and $\dot A,\dot B=1\ldots 2d$ are spinor 
and conjugate spinor indices of $SO(p-1,2)$.}
\be
g^A{}_B = \left(\begin{array}{cc}
L_\beta{}^\alpha+\frac{1}{2}D\delta_\beta{}^\alpha& -iK_{\alpha\dot\alpha} \\
-iP^{\dot\alpha \alpha} & -\bar L^{\dot\alpha}{}_{\dot\beta}-\frac{1}{2}D
\delta^{\dot\alpha}{}_{\dot\beta}\end{array}\right)
\label{eq:gpelems}
\ee
The $K$ and $P$ generate conformal transformations and conformal momentum, which are related
to $AdS$ conformal transformations and momenta by
\bea \tilde K &=& (K-P)/2 \nn \\
\tilde P &=& (K+P)/2 \ .
\eea
The $L_\beta{}^\alpha$ generate the Lorentz group and the $D$ are dilatations.
The stability group $H$ is generated by the $L$'s and the $\tilde K$'s.

First we must choose an $H$-invariant initial hypersurface. We can write 
this surface in the form
\be
\Z_0^A = \left(\begin{array}{c} \lambda_{0\alpha} \\ \mu_0^{\dot\alpha}
\end{array}\right)
\ee
As in the Penrose case, $L$-invariance requires that if $\Z_0$ contains any 
independent points with
$\lambda_0\not=0$, then it contains all such points, and similarly for
$\mu_0$. Since we want $0<\dim\Z_0 \le 2d-p$, only one of these two
should be independent of the other. Without loss of generality we 
choose $\lambda_0$ to be independent, and fix $\mu_0$ by
\be
\mu_0^{\dot\alpha} = F^{\dot\alpha\beta}\lambda_{0\beta} + G^{\dot\alpha
\dot\beta}\bar\lambda_{0\dot\beta}
\ee
for some $F^{\dot\alpha\beta}$ and $G^{\dot\alpha\dot\beta}$ which 
parametrize our twistorization. $\tilde K$-invariance then requires that
\bea
F\gamma_\mu F + G \gamma_\mu \bar G &=&  \gamma_\mu  \\
F\gamma_\mu G + G \gamma_\mu \bar F &=& 0
\label{eq:constrs}
\eea
where the $\gamma_\mu^{\alpha\dot\alpha}$ are
the Dirac matrices for $SO(p-2,1)$. For simplicity we will consider
the case $F=0$, so
\be
\Z_0^A = \left(\begin{array}{c}\lambda_{0\alpha} \\ G^{\dot\alpha\dot\beta}
\bar\lambda_{0\dot\beta} \end{array}\right)\ .
\ee

A simple choice of coset representative is 
\be
v(x^\mu,\rho) = e^{x\cdot P}e^{D\log\rho} = \left(\begin{array}{cc}
\rho^{1/2} & 0 \\ -i\rho^{1/2} x^{\dot\alpha\alpha} & \rho^{-1/2} 
\end{array}\right)\ ;
\ee
using this, and defining $\lambda=\rho^{1/2}\lambda_0$, the twistor is
\be
\Z^A(x^\mu,\rho) = \left(\begin{array}{c}\lambda_\alpha \\
-ix^{\dot\alpha\alpha}\lambda_\alpha + \rho^{-1}G^{\dot\alpha\dot\beta}
\bar\lambda_{\dot\beta} \end{array}\right)\ .
\label{eq:adstwistor}
\ee
As a check, the isometries of the space can be calculated from
\be
\delta\Z^A = g^A{}_B \Z^B\ .
\ee
Varying both sides of (\ref{eq:adstwistor}), one finds
\be
\delta\lambda_\alpha =\left(L_\alpha{}^\beta + \frac{1}{2}D \delta_\alpha
{}^\beta + K_{\alpha\dot\alpha} x^{\dot\alpha\beta} \right) \lambda_\beta
-i\rho^{-1} K_{\alpha\dot\alpha} G^{\dot\alpha\dot\beta}\bar\lambda_{\dot\beta}
\ee
and so
\bea
\delta x^{\dot\alpha\alpha} &=& P^{\dot\alpha\alpha} - x^{\dot\alpha\beta}
L_\beta{}^\alpha - \bar L^{\dot\alpha}{}_{\dot\beta} x^{\dot\beta\alpha}
+ D x^{\dot\alpha\alpha} + x^{\dot\alpha\beta}K_{\beta\dot\beta}
x^{\dot\beta\alpha} - \rho^{-2}K^{\dot\alpha\alpha} \\
\delta \rho &=& -D \rho - 2 \rho x\cdot K
\eea
which are the well-known isometries of anti-de Sitter space.

Geometric invariants may now be constructed by contracting $\Z$ with the
$SO(p-2,1)$ metric
\be
H_{\dot A}{}^B = \left(\begin{array}{cc}0&\C^{\dot\alpha\dot\beta} \\ \bar\C^{\dot\alpha
\dot\beta} & 0 \end{array}\right)
\ee
where $\C$ is the charge conjugation matrix, so
\be
\bar\Z_1 \cdot \Z_2 = \bar\lambda_1 \mu_2 + \bar\mu_1 \lambda_2\ .
\ee
A natural first guess for a particle action is
\be
\L = i\bar\Z \partial \Z\ .
\label{eq:twistbareaction}
\ee
This matches the kinetic term in (\ref{eq:cosetaction})
if we identify components of $\lambda$ with the momenta as follows:
\bea
P_{\alpha\dot\alpha} &=& 2\lambda_\alpha\bar\lambda_{\dot\alpha} \nn \\
P_\rho &=& \frac{i}{2\rho^2} \left[\bar\lambda G \bar\lambda - \lambda
\bar G \lambda \right] \ .
\label{eq:momentdef}
\eea

In (\ref{eq:twistbareaction}), however, all the components of $\lambda$ are 
independent and so their dynamics must be specified.  The first condition 
is the mass-shell constraint,
\be
\frac{1}{2\rho^2}P^2 - \rho^2 P_\rho^2 = \frac{1}{4}\left(\bar\Z\Z\right)^2
= M^2 R^2\ .
\label{eq:massshell}
\ee
There exist further independent components of $\lambda$ for most values
of $p$. These may be fixed by fixing the values of a set of twistor
bilinears
\be
\phi_i \equiv \bar\Z T_i \Z
\ee
where $(T_i)_{\dot A}{}^B$ are some constant matrices which transform 
in the $(\frac{1}{2},\frac{1}{2})$ of $SO(p-1,2)$.
The number of independent $\phi_i$ that must be set depends on $p$. In an
action, these will be constrained to values $m_i$. For example, using
(\ref{eq:massshell}) the mass-shell constraint is $\phi_{T=1}=2MR$.
So the complete twistor action takes the simple form
\be
\L = i\bar\Z\left(\partial -iu^i T_i\right)\Z - u^i m_i
\label{eq:twistaction}
\ee
where the $u^i$ are Lagrange multipliers. This action is equivalent to
(\ref{eq:cosetaction}). It has several important features:

\begin{enumerate}
\item{The action is manifestly $SO(p-1,2)$ invariant and has a quadratic
kinetic term. It has the structure of a world-line gauge theory with
sources. The ``gauge fields'' $u^i$ are nondynamical since there is no
field strength in one dimension. 

This statement can be made somewhat more precise by noting that
 (\ref{eq:twistbareaction}) implies that the Poisson brackets (which
will become commutators in the quantized theory) are
\be
\left\{\Z_A, \bar\Z^{\dot B}\right\}_{PB} = -2i H_A{}^{\dot B}
\ee
with all other brackets vanishing, and so
\be
\left\{\phi_i, \phi_j\right\}_{PB} = -2i\bar\Z\left[T_i, T_j\right]\Z\ .
\ee
Since the set of constraints under Poisson brackets forms a Lie algebra,
the set of $T_i$ form one as well, and this algebra is invariant under 
$SO(p-1,2)$. This guarantees that the action (\ref{eq:twistaction}) indeed
has a gauge symmetry.
}
\item{The gauge group contains a $U(1)$ factor corresponding
to the mass-shell constraint $T=1$, $m=2MR$. The rest of the  group may 
be calculated
explicitly for small $p$ by constructing the $\phi_i$; they are

\begin{center}
\begin{tabular}{c|*{7}{c}}
p&1&2&3&4&5&6&7 \\ \hline
$\dim\Z_0$ & 2&2&4&4&8&8&16 \\
$N_\phi$ & 1 & 0 & 1 & 0 & 3 & 2 & 9 \\
Group & $U(1)^2$ & $U(1)$ & $U(1)^2$ & $U(1)$ & $U(1)\times SU(2)$
& $U(1)^3$? & $U(1)\times SU(2)^3$?
\end{tabular}
\end{center}

The final two are conjectured but have not been explicitly calculated.

This is related to the result of \cite{CRZ,CKR} for $AdS_5$. In that
case, the 8-component spinors were decomposed into a pair of 4-component
spinors of the stability group $H=SO(4,1)$ indexed by $I,J=1,2$, and
\be
(T_i)_{aI}{}^{bJ} = (\sigma_i)_I{}^J \C_a{}^b
\ee
(The $a,b$ are $SO(4,1)$ spinor indices)
}
\item{This twistor Lagrangian can be quantized following a procedure
similar to that used in \cite{CGKRZ}, leading to solutions which transform
in representations of $SO(p-1,2)$.}
\item{For $i\not=0$, The $\phi_i$ may be chosen to be independent 
of the momenta. In these cases it is not clear what meaning one
could assign to a nonzero $m_i$. The analogous quantities in 
\cite{CRZ,CKR} are all zero.}
\item{The Lagrange multipliers $u^i$ can be integrated out to give
\be
\L'(k) = i\left.\bar\Z(\partial+iT\cdot m)\Z\right|_k + \int
\frac{dq}{2\pi}\left.\left(\bar\Z T^i\Z\right)\right|_{k+q}
\left.\left(\bar\Z T_i \Z\right)\right|_{k-q}
\label{eq:twistaction2}
\ee
which is therefore equivalent to (\ref{eq:twistaction}). (This can also
be seen by explicitly resumming Feynman diagrams involving the $u^i$)
}
\end{enumerate}

\bigskip

The actions (\ref{eq:twistaction}) and (\ref{eq:twistaction2}) represent
a considerable simplification over their classical counterpart
(\ref{eq:cosetaction}). Because they have leading quadratic terms and
manifest $G$-symmetry, their quantum solutions automatically fill out
representations of the isometry group. A similar construction can be
carried out for an arbitrary coset manifold, or even a supercoset, and
(similarly to \cite{CGKRZ}) can be used to construct string actions on
these spaces. Since the known superstring actions are manifestly invariant
under the isometries, it is likely that  these systems will be amenable
to a twistor interpretation which would allow their quantization and
analysis, including interactions with Ramond-Ramond and Neveu-Schwarz
background fields.

\bigskip
\centerline{{\bf Acknowledgements}}

The author wishes to thank Piet Claus, Renata Kallosh, Michael Peskin,
Joachim Rahmfeld, and Steve Shenker
for numerous useful discussions and comments, and the organizers of the 
TASI-1999
school where part of this work was accomplished. This work was supported
in part by an NSF Graduate Research Fellowship and by NSF grant
PHY-9870115.


\begin{thebibliography}{99}
\bibitem{Malda} J.~Maldacena,
\newblock ``The Large N limit of superconformal field theories and
supergravity,'' Adv. Theor. Math. Phys. {\bf 2} (1998) 231,
hep-th/9711200
\bibitem{GubKleb}
S.~Gubser, I.~R.~Klebanov and A.~M.~Polyakov, ``Gauge theory
correlators from noncritical string theory,'' Phys. Lett. {\bf B428}
(1998) 105, hep-th/9802109
\bibitem{Witten}E.~Witten, ``Anti-de Sitter space and holography,'' Adv. Theor.
Math. Phys. {\bf 2} (1998) 253, hep-th/9802150.
\bibitem{Penrose} R.~Penrose and W.~Rindler, {\em Spinors and Space-time}
vol. 2, section 6.3; Cambridge U.P., Cambridge, 1986
\bibitem{Ferber} A.~Ferber, ``Supertwistors and Conformal
Supersymmetry,'' Nucl. Phys. {\bf B132} (1978) 55.
\bibitem{CRZ} P.~Claus, J.~Rahmfeld and Y.~Zunger,
``A Simple particle action from a twistor parametrization of AdS(5),''
hep-th/9906118.
\bibitem{CKR} P.~Claus, R.~Kallosh and J.~Rahmfeld,
``BRST quantization of a particle in AdS(5),''
hep-th/9906195.
\bibitem{BlueBook} L.~Castellani, R.~D'Auria, and P.~Fr\'e, {\em Supergravity 
and superstrings : a geometric perspective}, \S 1.6; World Scientific 
Press, Teaneck, N.J. (1991)
\bibitem{Townsend} P.~K.~Townsend, ``Supertwistor Formulation of the
Spinning Particle,'' Phys. Lett. {\bf B261} (1991) 65.
\bibitem{CGKRZ}P.~Claus, M.~Gunaydin, R.~Kallosh, J.~Rahmfeld and Y.~Zunger,
``Supertwistors as quarks of SU(2, 2$|$4),''
JHEP {\bf 05}, 019 (1999)
hep-th/9905112.
\end{thebibliography}
\end{document}